%% file: proc.tex
\documentclass[cits]{PoS}

\usepackage{epsfig}
\usepackage{graphicx}

\include{macros}

\title{Chiral behavior of pseudo-Goldstone boson masses and decay
  constants in $2+1$ flavor QCD}

\ShortTitle{Chiral behavior of pseudo-Goldstone boson masses and decay
  constants}

\author{S.~D\"urr$^{1}$, Z.~Fodor$^{1,2,3}$, C.~Hoelbling$^{2,5}$,
S.D.~Katz$^{2,3}$,
S.~Krieg$^{2,4}$, Th.~Kurth$^2$, \speaker{L.~Lellouch}$^{,5}$,
Th.~Lippert$^{1,2,4}$, K.K.~Szabo$^2$, G.~Vulvert$^5$
\\ \\
$^1$John von Neumann Institute for Computing (NIC), DESY, D-15738, Zeuthen
$/$ 
FZJ, D-52425, Juelich, Germany
\\
$^2$Department of Physics, University of Wuppertal, D-42097 Wuppertal,
Germany 
\\
$^3$Institute for Theoretical Physics, E\"otv\"os University, H-1117
Budapest, Hungary 
\\
$^4$J\"ulich Supercomputing Center (JSC), Forschungszentrum J\"ulich,
D-52425 J\"ulich, Germany
\\
$^5$Centre de Physique Th\'eorique, Case 907, Campus de Luminy,
    F-13288 Marseille Cedex 9, France\thanks{CPT is ``UMR 6207 du CNRS
    et des universit\'es d'Aix-Marseille I, d'Aix-Marseille II et du
    Sud Toulon-Var, affili\'ee \`a la FRUMAM.''}

\\ \\ E-mail: \email{lellouch@cpt.univ-mrs.fr}}

\abstract{We present preliminary results for the chiral behavior of
charged pseudo-Goldstone-boson masses and decay constants. These are
obtained in simulations with $N_f{=}2{+}1$ flavors of tree-level,
$O(a)$-improved Wilson sea quarks. In these simulations, mesons are
composed of either valence quarks discretized in the same way as the
sea quarks (unitary simulations) or of overlap valence quarks
(mixed-action simulations). We find that the chiral behavior of the
pseudoscalar meson masses in the mixed-action calculations cannot be
explained with continuum, partially-quenched chiral perturbation
theory. We show that the inclusion of $O(a^2)$ unitarity violations in
the chiral expansion resolves this discrepancy and that the size of
the unitarity violations required are consistent with those which we
observe in the zero-momentum, scalar-isotriplet-meson propagator.}

\FullConference{The XXV International Symposium on Lattice Field Theory\\
		 July 30 - August 4 2007\\
		 Regensburg, Germany}

\begin{document}

\section{Introduction}

The objective of our collaboration is to calculate hadronic
observables which are relevant for determining fundamental quark
properties, such as quark masses or quark flavor-mixing and
CP-violation parameters, and to do so with controlled extrapolations
to the physical limit of $N_f{=}2{+}1$ flavor QCD, where $M_\pi\simeq
135\,\mev$, the lattice spacing $a$ vanishes and the volume is
infinite. To achieve that goal we consider two approaches. In both,
the seas are composed of $N_f{=}2{+}1$ flavors of tree-level,
$O(a)$-improved Wilson (W) quarks~\cite{Sheikholeslami:1985ij}. We
perform ``unitary'' simulations where the valence quarks are
discretized in the same way as the sea quarks. We also perform
``mixed-action'' calculations, with
overlap~\cite{Kaplan:1992bt,Narayanan:1995gw,Neuberger:1998fp},
Ginsparg-Wilson (GW)~\cite{Ginsparg:1982bj} valence quarks. In the
latter, the valence sector possesses a full, continuum-like chiral
symmetry~\cite{Luscher:1998pq} which greatly simplifies the
renormalization of electroweak operators, such as those encountered in
neutral kaon mixing. It also guarantees that matrix elements are
automatically $O(a)$-improved, to the extent that the sea quarks are.
As discussed by Stefan Krieg at this conference~\cite{stefan}, recent
advances have allowed us to perform $N_f{=}2{+}1$ simulations, for
instance, down to $M_\pi\sim 190\,\mev$ with $a\sim 0.09\,\fm$ and in
cubic volumes of side $L \sim 4.2\,\fm$. Thus, we expect to be able to
reach the near-continuum chiral $p$-regime of Gasser and Leutwyler
without the conceptual problems of staggered
fermions~\cite{Sharpe:2006re,Creutz:2007rk}. This means that we should
be able to extrapolate lattice results to the physical point in a
model-independent way, by using
Wilson~\cite{Sharpe:1998xm,Rupak:2002sm,Aoki:2003yv,Bar:2003mh},
partially quenched (PQ) chiral perturbation theory
($\chi$PT)~\cite{Sharpe:1992ft,Bernard:1992mk,Bernard:1993sv,Sharpe:2001fh}
for the unitary simulations, and mixed-action (MA)
PQ$\chi$PT~\cite{Bar:2002nr,Bar:2003mh,Chen:2007ug} for the GW-on-W
simulations.

One of the drawbacks of using a mixed-action approach is the
presence of discretization-induced unitarity violations. Fortunately,
it should be possible to account for the low-energy manifestations of
these violations with MAPQ$\chi$PT. We present here preliminary
results for the quark-mass dependence of the masses and leptonic decay
constants of the pseudo-Goldstone bosons (PGBs) of chiral symmetry
breaking. In particular, we investigate the effects of unitarity
violations in these quantities, as obtained in our mixed-action
simulations, and attempt to correlate these effects with those which
we observe in the scalar, isotriplet, $a_0$ propagator, where they are
expected to be particularly large \cite{Golterman:2005xa}.  Of course,
our study of the masses and decay constants of the PGBs is primarily
motivated by the very interesting phenomenology they give rise
to. They allow the determination of a variety of fundamental
quantities, such as light quark masses, the ratio of CKM elements
$|V_{us}/V_{ud}|$~\cite{Marciano:2004uf} and important LO and
higher-order low-energy constants (LECs) of the effective chiral
Lagrangian. However, we postpone the presentation of results for
these quantities to later publications.

\section{Finite-volume mixed-action PQ$\chi$PT and unitarity violations}
\label{sec:unitarityviolations}

As shown in \cite{Golterman:2005xa} for the case of $N_f$ degenerate
flavors, the propagator of the $a_0$ is affected by potentially large
$O(a^2)$ unitarity violations in a mixed-action scenario. For the case
which interests us, with $N_f{=}2{+}1$, to simplify expressions we suppose
that the light sea ($\ell$) and valence ($v$) quark masses are tuned
such that masses of the corresponding PGBs are equal:
$M_{vv}=M_{\ell\ell}\stackrel{\cdot}{\equiv} M_\pi$. We denote the
strange sea quark by $s$, and by $M_{ss}$ the mass of charge
pseudoscalar mesons composed of two valence quarks with the mass of
the $s$.~\footnote{From now on we call these mesons ``non-singlet
$s\bar s$'' pseudoscalar mesons.} Then, LO MAPQ$\chi$PT gives for the
zero-momentum, $a_0$ propagator:
\bea C_{a_0}(t,\lambda) &\equiv &Z_S^2(g_0,a\lambda)\,a^3\sum_{\vec{x}}
\langle \bar q_2 q_1(\vec{x},t)\bar q_1
q_2(0)\rangle\label{eq:a0prop}\\ &\stackrel{t\to
+\infty}{\longrightarrow}& \frac{B^2(\lambda)}{L^3}\left\{C_{K\bar
K}(t)+ \frac23 C_{\pi\eta}(t)-2\frac{a^2\Delta}{M_\pi^2}\left( M_\pi\,
t+1\right)C_{\pi\pi}(t)\right\}\nn \ ,\eea
where the valence quarks are chosen to be degenerate, with
$m_1=m_2\stackrel{\cdot}{\equiv}m_v$, and where we have assumed, for
simplicity, that the time and space extent of the lattice are
infinite. The quantity $B$ is the condensate over the square of the
decay constant $F$. $Z_S(g_0,a\lambda)$ is a renormalization constant
for scalar densities and $\lambda$ is a QCD renormalization
scale. The functions $C_{XY}(t)$ denote the zero-momentum propagators
of the two-particle states, $XY$, and are given by
$C_{XY}(t)\equiv\exp\bigl[-(M_X+M_Y)t\bigr]/(4M_XM_Y)$ at LO. There
are two physical contributions, coming from intermediate two-kaon and
$\pi\eta$ states. The unitarity violations are of order $a^2$ and are
proportional to a quantity $\Delta$, which has mass dimension
four. These violations only vanish in the continuum limit. Moreover,
they are exponentially and polynomially enhanced in $t$, also compared
to the contribution from the $a_0$ not shown in \eq{eq:a0prop}. At
asymptotic times, the unitarity violations are the dominant
contribution.

Unitarity violations also affect PGB masses and decay
constants, but only at NLO.  Let us consider a pseudoscalar meson
composed of two distinct quarks with masses $m_1$ and $m_2$. Then,
according to NLO MAPQ$\chi$PT, the square of this meson mass has the
following generic form:
\be
\begin{array}{rcl}
(M_{12}^2)^{\NLO}_\Omega&=&(m_1+m_2)B\Bigl\{1+\frac1{(4\pi F)^2}
\Bigl[\mbox{PQ-logs$(\mu,M_{11},M_{22},M_{\ell\ell},M_{ss})$}\\[0.2cm]
&&\qquad\qquad + (2\alpha_6-\alpha_4)(\mu)(2M_{\ell\ell}^2+ M_{ss}^2)
+ (2\alpha_8-\alpha_5)(\mu)M_{12}^2+
\mbox{FV}\\[0.2cm]
&&\qquad\qquad+a\beta_M+a^2\Delta\times\{
\mbox{UV-logs$(\mu,M_{11},M_{22})$}+\gamma_M(\mu)\}\Bigr]\Bigr\}\ ,
\end{array}
\label{eq:nloM122}
\ee
where $B$ and $\Delta$ are defined after \eq{eq:a0prop}; ``PQ-logs''
and ``UV-logs'' denote partially-quenched and unitarity-violating
quenched-like logarithms, respectively; $\mu$ is the renormalization
scale in the effective theory and, in \eq{eq:nloM122}, the quark
masses and the $B$ must both be either renormalized in the same
scheme in QCD or bare; $\gamma_M(\mu)$ is an a priori unknown
counter-term; the LECs $\alpha_i$ are related to the original Gasser
and Leutwyler constants through $\alpha_i(\mu)\equiv
8(4\pi)^2L_i(\mu)$; FV stands for finite-volume corrections; $\beta_M$
is a mass-dimension three quantity which parameterizes $O(a)$
discretization errors and whose parametric size will be specified
below. Note that in fits to lattice results obtained at a single
lattice spacing, the discretization errors proportional to $a\beta_M$
and $a^2\Delta\gamma_M(\mu)$ get absorbed into the LO LEC, $B$.

In applying the general form of \eq{eq:nloM122}, it is useful to
distinguish three cases of interest. In the continuum or in the case
of GW valence on GW sea quarks, $m_1$ and $m_2$ can be taken to be the
Lagrangian masses and $\Delta$ and $\beta_M$ identically vanish,
i.e. there are no unitarity violating nor $O(a)$ discretization
errors. In our W-on-W unitary simulations, we take $m_1$ and $m_2$ to
be the ``measured'' axial Ward identity (AWI) masses. Moreover, the
constant $\beta_M$ is $O(\lqcd^3)$ if the fermions used are straight
Wilson fermions, $O(\alpha_s\lqcd^3)$ if they are tree-level
$O(a)$-improved as they are in our simulations, or zero if they are
non-perturbatively $O(a)$-improved. Finally, $\Delta\equiv 0$ and
discretization-induced unitarity violations are absent.  In the
mixed-action, GW-on-W case, $m_1$ and $m_2$ can be taken to be the
valence, overlap Lagrangian masses. $\beta_M$ has the same parametric
size as for the unitary simulations, depending on the level of
improvement of the sea. However, in the mixed-action case, $\Delta$
does not vanish a priori. Thus, we expect that meson masses will
suffer from discretization-induced unitarity violations at finite $a$.

The NLO expression for the decay constant, $F_{12}$, of a charged
pseudoscalar meson is similar in form to \eq{eq:nloM122}, with the
factor $(m_1+m_2)B$ replaced by $F$ and where the ``PQ-logs''and
corresponding counter-terms and finite-volume corrections are changed
to the partially-quenched expressions appropriate for decay
constants. The corresponding general form applies to the three cases
discussed above in much the same way, with the substitutions
$\beta_M\to\beta_F$ and $\gamma_M\to\gamma_F$. The main difference
appears in the GW-on-W case, where the mixed-action unitarity
violations proportional to $a^2\Delta$ are purely valence
$SU(3)$-flavor breaking, and do not depend on $\mu$.

\section{Results for the $a_0$ propagator and the charged
  PGB masses and decay constants}

The results discussed below were obtained from simulations which were
described at this conference by Stefan Krieg~\cite{stefan}. We
recapitulate here only the main ingredients, omitting algorithmic
considerations. The gauge action used is tree-level Symanzik
improved~\cite{Luscher:1984xn}. The sea quarks are described by
six-step stout-smeared~\cite{Morningstar:2003gk}, tree-level
$O(a)$-improved Wilson fermions~\cite{Sheikholeslami:1985ij}. For the
unitary, W-on-W simulations, the valence quarks are discretized in the
same way. In the GW-on-W, mixed-action case, valence quarks are
three-step HYP-smeared~\cite{Hasenfratz:2001hp} overlap
fermions~\cite{Kaplan:1992bt,Narayanan:1995gw,Neuberger:1998fp}, with
a negative mass parameter $\rho=1$.

We have performed five, $2{+}1$ sea-flavor simulations at a lattice
spacing $a\sim 0.09\,\fm$ ($\beta=3.57$). In these simulations, the
mass of the charged pions composed of two light sea quarks are $M_\pi\sim
190$, 300, 410, 490 and $570\,\mev$. To keep finite-volume errors
small, all simulations are performed in cubic three-volumes with sides
$L$ such that $M_\pi L\ge 4$. The strange quark mass used in our
simulations is slightly overestimated: with such a strange quark, the
mass of a kaon, when extrapolated in light-quark mass to the physical
point, is approximately $7\%$ higher than the physical kaon
mass. There are 34 gauge configurations at $M_\pi\sim 190\,\mev$, 68
at $M_\pi\sim 300\,\mev$ and $O(100)$ at the three other simulation
points. In the mixed-action case, the overlap quark masses are chosen
such that the mesons which they compose are approximately degenerate
with those composed of the corresponding Wilson sea quarks. At
$M_\pi\sim 190$ and $300\,\mev$, we have a second overlap valence
strange quark whose mass is approximately 30\% smaller than that of
the strange sea quark.

\begin{figure}
\begin{center}
\psfig{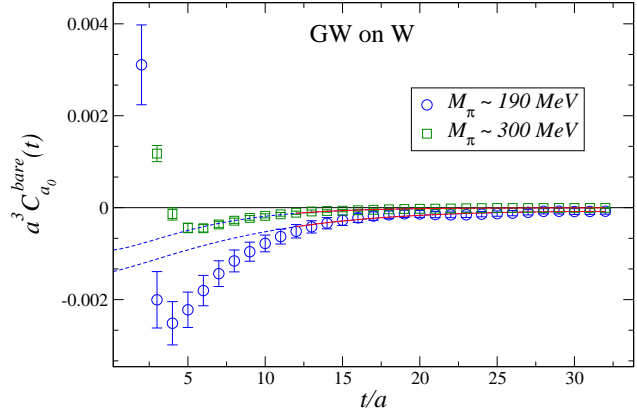}
\caption{\label{fig:a0prop} Bare, zero-momentum propagator of the $a_0$ as a
function of time over half the time extent of our lattices, as
obtained in our two GW-on-W simulations with the lightest $u$ and $d$
quarks. The solid curves represent our best fits to the
partially-quenched and finite-volume generalization of
\protect\eq{eq:a0prop} in the fit region, and the dashed curves their
extensions to earlier times.}
\end{center}
\end{figure}

We begin with preliminary results for the zero-momentum, $a_0$
propagator, $C_{a_0}(t)$, defined in \eq{eq:a0prop}. In
\fig{fig:a0prop} we plot the unrenormalized propagators,
$C_{a_0}^{\,\bare}(t)$, as a function of Euclidean time $t$, obtained in
the two GW-on-W simulations with the lightest $u$ and $d$ quarks. The
propagators go negative at relatively short times and then
asymptotically go back up to zero. Moreover, the effect is less
pronounced for the simulation with the more massive quarks, a trend
which persists as one increases the $u$ and $d$ quark masses
further. This behavior is qualitatively consistent with the prediction
of MAPQ$\chi$PT given in \eq{eq:a0prop}, assuming $\Delta>0$.  The
agreement can be made quantitative also. To verify this, we perform
fits of the propagators, at asymptotic times $t$, to the PQ and FV
generalization of \eq{eq:a0prop} for the bare $C_{a_0}(t)$. The fits
have only one parameter, namely $a^4\Delta$. For the pre-factor
$(B/Z_S)^2$, we take $M_{12}^2/(m_1+m_2)$, where $M_{12}$ is the
``measured'' meson mass and $m_{1,2}$ are the bare masses of the GW
quarks which compose it. The kaon, pion and $\eta$ masses which appear
in the expression for $C_{a_0}^{\,\bare}(t)$ are constrained to take on
the values obtained from prior fits to kaon and pion two-point
functions.  These mesons are composed of a sea and a valence quark. We
obtain their masses by combining the corresponding valence-valence and
sea-sea masses at LO in the chiral expansion. This means that the
parameter $a^4\Delta$ that we fit here contains contributions from an
$O(a^2)$, mixed-action operator in the chiral Lagrangian. Fortunately,
if the same prescription is used for obtaining valence-sea meson
masses which appear at NLO in quantities such as $M_{12}^2$ or
$F_{12}$, the unitarity violations there will be parameterized by the
same $a^4\Delta$.

The one-parameter fits to $C_{a_0}^{\,\bare}(t)$ for $M_\pi\sim 190$ and
$300\,\mev$ are performed for $t/a$ in the range $[12,32]$, where
$t/a=32$ is the midpoint of our lattices in both cases. The results of
these fits are plotted in \fig{fig:a0prop}. As the figure suggests,
both fits have good $\chi^2/dof$. The values obtained for the
unitarity-violation parameter are $a^4\Delta=0.015(6)$ and
$0.024(10)$, respectively, for $M_\pi\sim 190$ and $300\,\gev$, and
are thus consistent. For the lattice spacing at which the simulations
are performed, these values correspond to $a\sqrt\Delta\sim
0.27\,\gev$ and $0.35\,\gev$. Since $a\sqrt\Delta$ competes with pion,
kaon and $\eta$ masses in the chiral expressions for $M_{12}^2$ and
$F_{12}$ in the mixed-action case, it is clear that these
unitarity-violating contributions cannot be neglected a priori.

We now turn to an analysis of the GW-on-W decay constant. We begin
with this quantity, because we use the extrapolated $aF_\pi$ to
determine the lattice spacing as well as to normalize corrections in
chiral expressions with factors of $(4\pi aF_\pi)^2$. $aF_{12}$ is
obtained from the pseudoscalar two-point function using the
AWI. Thanks to the chiral symmetry of the overlap, valence quarks, no
renormalization is required. This is a simple example of the
simplifications brought about by the use of a mixed action with
chirally symmetric valence quarks.

There are 21 lattice points for $aF_{12}$, of which 5 correspond to
charged ``pions'', 7 to ``kaons'' and 9 to ``non-singlet $s\bar s$''
pseudoscalar mesons. We fit these results to the NLO chiral expression
described in \sec{sec:unitarityviolations}. So as to remain, as much
as possible, within in the range of applicability of NLO $\chi$PT, we
include in the fit only the 4 lightest pion and 4 lightest kaon
points, with $M_\pi\le 500\,\mev$ and $M_K\le 590\,\mev$. The fit has
four parameters, which are $F$, $\alpha_4(M_\eta)$, $\alpha_5(M_\eta)$
and $a^4\Delta$. Since the $a_0$ propagator is more sensitive to
unitarity violations than are the decay constants, we fix $a^4\Delta$
to the value 0.024(10), through a Gaussian prior in the $\chi^2$. The
NLO expression describes the data well. Moreover, the resulting value
of $a^4\Delta$ is 0.025(8), confirming that the chiral behavior of the
decay constants is consistent with the presence of unitarity
violations of the size observed in the $a_0$ propagator.  The value of
$aF_\pi$ obtained from a self-consistent extrapolation to the physical
point, using the infinite-volume, continuum fitted function, yields a
lattice spacing of $a=0.088(1)\,\fm$, where the error is statistical.

\begin{figure}
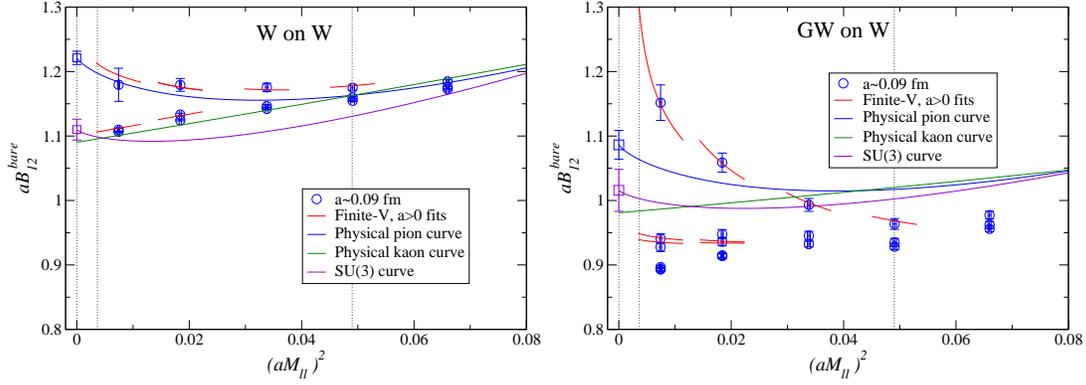

\begin{center}
\psfig{file=Fit_aB12_vs_aMll_sq_LT600MeV_FVDE_NLO_nb200_15pts_WonW.eps,width=0.47\textwidth,angle=0}
\psfig{file=Fit_aB12_vs_aMll_sq_LT600MeV_FVDE_NLO_nb200_21pts.eps,width=0.47\textwidth,angle=0}
\caption{\label{fig:aB12vsaMll2WonW} The bare condensate ratio as a
function of squared sea-pion mass, in lattice units, for the unitary
simulations (left) and for the mixed-action simulations (right). The
vertical scales in the two plots are only equal up to a ratio of
renormalization constants. The three sets of data (circles) in the
W-on-W case correspond, from top to bottom, to pion-like, kaon-like
and non-singlet $s\bar s$-like pseudoscalar mesons. In the GW-on-W
plot, there are one additional kaon and two extra $s\bar s$ points at
the two smallest $M_{\ell\ell}$'s, corresponding to the additional
valence strange quark that we consider in these simulations. The fits
to \protect\eq{eq:nloM122} are plotted as line-segments around each
fitted point. The physical curves are obtained from the fits by
removing the FV effects, and in the mixed-action case, the
partial-quenching effects and unitarity-violating logs. The pion
curves are obtained by setting $M_{11}{=}M_{22}{=}M_{\ell\ell}$ and
$M_{ss}{=}M_{ss}^\mathrm{phys}$, the latter being the physical,
non-singlet $s\bar s$ pseudoscalar meson mass; the kaon curves, by
setting $M_{11}{=}M_{\ell\ell}$ and
$M_{22}{=}M_{ss}{=}M_{ss}^\mathrm{phys}$; the $SU(3)$ curves, by
setting $M_{11}{=}M_{22}{=}M_{ss}{=}M_{\ell\ell}$. The vertical dotted
lines mark, from left to right, the chiral limit, the physical pion
and the physical kaon points.}
\end{center}
\end{figure}

Next we consider chiral fits to the unitary W-on-W results for the
condensate ratio $aB_{12}^{\,\bare}\equiv
(aM_{12})^2/(am_1+am_2)^\bare_\AWI$. The lattice results for this
quantity are plotted in \fig{fig:aB12vsaMll2WonW}, as a function of the
squared sea-pion mass, $M_{\ell\ell}^2$. The different sets of points
correspond to ``pions'', ``kaons'' and ``non-singlet $s\bar s$''
pseudoscalar mesons. We fit these results to the NLO expression of
\eq{eq:nloM122} with $\Delta\equiv 0$. We include in the fit only the
six points with $M_\pi\le 500\,\mev$ and $M_K\le 590\,\mev$. Here, the
fit has only three parameters, namely $B$,
$(2\alpha_6-\alpha_4)(M_\eta)$ and $(2\alpha_8-\alpha_5)(M_\eta)$. The
NLO expression describes the chiral behavior of $aB_{12}^{\,\bare}$ very
well.

We now turn to the condensate ratio obtained in the mixed-action,
GW-on-W simulations. Here, $aB_{12}^{\,\bare}\equiv
(aM_{12})^2/(am_1+am_2)^\bare$, where $M_{12}$ is the valence meson
mass and $m_{1,2}^\bare$ are the corresponding bare overlap Lagrangian
masses. The results for this quantity are plotted in
\fig{fig:aB12vsaMll2WonW}, again as a function of $M_{\ell\ell}^2$. As
the plot indicates, the behavior of $aB_{12}^{\,\bare}$ here deviates
significantly from that obtained in the unitary case. Moreover, some
of the features of this behavior, such as the large increase of
$aB_{12}^{\,\bare}$ for the ``pion'' points at small $M_{\ell\ell}^2$,
cannot be explained with only continuum PQ chiral logarithms: a
divergent term at small $M_{\ell\ell}^2$ appears to be
required. Fortunately, such a contribution is provided by the
unitarity violations discussed in \sec{sec:unitarityviolations} and
exhibited in \eq{eq:nloM122}. We thus fit the lattice results to the
NLO expression of \eq{eq:nloM122}, including the unitarity violating
term proportional to $a^2\Delta$. As for the decay constant fit, we
fix $a^4\Delta$ to the value 0.024(10), obtained from the $a_0$
propagator, through a Gaussian prior in the $\chi^2$. There are four
parameters in the fit, one more than in the W-on-W case. These are
$B$, $(2\alpha_6-\alpha_4)(M_\eta)$, $(2\alpha_8-\alpha_5)(M_\eta)$
and a constrained $a^4\Delta$. Again, only the eight points with
$M_\pi\le 500\,\mev$ and $M_K\le 590\,\mev$ are included. The
description of the condensate ratio given by our NLO chiral expression
is good. The value of $a^4\Delta$ returned by the fit is 0.020(6). As
already noted after \eq{eq:nloM122}, in a fit performed at fixed
lattice spacing, the discretization error proportional to
$\gamma_M(\mu)$ gets absorbed into $B$, which thereby acquires a
spurious $\chi$PT $\mu$-dependence. Of course, this dependence will be
eliminated, along with all other discretization errors, when the
renormalized values of $B$, obtained at different lattice spacings,
are extrapolated to the continuum limit. The value of $\mu$ chosen
here is $M_\eta$. A lower value will raise the physical curves whereas
a larger one will lower them. It is worth noting that this spurious
$\mu$-dependence cancels at NLO in ratios such as $m_s/m_{ud}$ or
$\langle\bar qq\rangle_{N_f{=}2}/\langle\bar
qq\rangle_{N_f{=}3}$. Moreover, it only affects the fitted LECs
$(2\alpha_6-\alpha_4)(M_\eta)$ and $(2\alpha_8-\alpha_5)(M_\eta)$ very
mildly. We find very good agreement between the GW-on-W and W-on-W results
for these quantities.

These observations, together with the other results reported on here,
suggest that unitarity violations are present in our mixed-action
results, and that we can subtract them with MAPQ$\chi$PT. Of course,
simulations at other lattice spacings are required to confirm this
conclusion.

\vspace{0.4cm}
\noindent
{\bf Acknowledgments}

Computations are performed on the BlueGene/L at FZ J\"ulich and on
clusters at the University of Wuppertal and at CPT Marseille. This work
is supported in part by EU grant I3HP, OTKA grant AT049652, DFG grant FO
502/1, EU RTN contract MRTN-CT-2006-035482 (FLAVIAnet) and by the CNRS's GDR
grant n$^\mathrm{o}$2921 (``Physique subatomique et calculs sur
r\'eseau'').

\bibliographystyle{my-elsevier}

\bibliography{proc}


\end{document}

%% file: macros.tex
\def\be{\begin{equation}}
\def\ee{\end{equation}}
\def\bea{\begin{eqnarray}}
\def\eea{\end{eqnarray}}

\def\reff#1{\ref{#1}}

\def\eq#1{Eq.~(\reff{#1})}

\def\fig#1{Fig.~\reff{#1}}

\def\sec#1{Sec.~\reff{#1}}

\def\nn{\nonumber}

\def\mev{\mathrm{Me\kern-0.1em V}}
\def\gev{\mathrm{Ge\kern-0.1em V}}
\def\fm{\mathrm{fm}}

\def\NLO{\mathrm{NLO}}
\def\bare{\mathrm{bare}}
\def\AWI{\mathrm{AWI}}

\def\lqcd{\Lambda_\mathrm{QCD}}

\newcommand{\lsim}{ {\
\lower-1.2pt\vbox{\hbox{\rlap{$<$}\lower5pt\vbox{\hbox{$\sim$}}}}\ } }
\newcommand{\gsim}{ {\
\lower-1.2pt\vbox{\hbox{\rlap{$>$}\lower5pt\vbox{\hbox{$\sim$}}}}\ } }